\documentstyle{article}\begin{document}
\title{Chern-Simons states of photons and gravitons}

\author{ Z. Haba\\
Institute of Theoretical Physics, University of Wroclaw,\\ 50-204
Wroclaw, Plac Maxa Borna 9, Poland.\\
email:zbigniew.haba@uwr.edu.pl} \maketitle
\begin{abstract} We consider an alternative quantization of the
electromagnetic field around the Chern-Simons state $\psi_{CS}$
which is a zero energy solution of quantum electrodynamics. The
solution determines a stochastic process which is a random
perturbation of the self-duality equation for the electromagnetic
potential ${\bf A}$. The stochastic process defines a solution of
the Schr\"odinger equation with the initial condition
$\psi_{CS}\chi$ where $\chi({\bf A})$ is an analytic function of
${\bf A}$ decaying fast at large ${\bf A}$. The method can be
applied to a quantization of non-Abelian gauge theories by means
of a stochastic self-duality equation. A quantization  of massless
spin 2 tensor fields (based on self-duality) and its extension to
a quantization of gravity are also discussed.

\end{abstract}

\section{Introduction}

We consider solutions of the Schr\"odinger equation with a
canonical Hamiltonian determined by the standard Lagrangian of the
electromagnetic field. We look for solutions in the WKB form
$\psi_{t}=\exp(\frac{i}{\hbar}W_{t})\chi_{t}$, where
$\exp(\frac{i}{\hbar}W_{t})$ is a particular solution of the
Schr\"odinger equation. In such a case the evolution of $\chi_{t}$
can be expressed by a random perturbation of the classical
solution determined by the Hamilton-Jacobi function $W$. We admit
complex $W$ and we do not require that
$\exp(\frac{i}{\hbar}W_{t})$ is square integrable. The
Chern-Simons wave functions $\exp(\frac{i}{\hbar}W_{CS})$ with the
Chern-Simons action $iW_{CS}$ (where $iW_{CS}$ is real) are
solutions of the Schr\"odinger equation. Such wave functions
appear  in quantum gravity \cite{kodama}. They are  rejected
\cite{witten} as unphysical ( being not square integrable ). We
show that such solutions of the Schr\"odinger equation acquire a
relevance in our approach to quantum theory. The method is based
on a particular solution $\exp(\frac{i}{\hbar}W_{t})$  of the
Schr\"odinger equation which determines a stochastic process which
defines a time evolution of the states of the form
$\psi=\exp(\frac{i}{\hbar}W)\chi$. $\psi$ is required to be square
integrable but $\exp(\frac{i}{\hbar}W)$ does not need to satisfy
this condition. It comes out that if $W$ is of the Chern-Simons
form then the randomly perturbed classical equation is a random
perturbation of the self-duality equation.

The appearance of the Chern-Simons terms in the effective
Lagrangians leads to interesting phenomena in condensed matter
physics in 2+1 dimensions (anyons \cite{wilczek}, charged vortices
with fractional statistics \cite{frohlich}) as well as in 2+1
\cite{wittensoluble} and 3+1 dimensional quantum gravity
\cite{polar0} \cite{polar1}\cite{polar2}. The wave functions of
the form $\exp(\frac{i}{\hbar}W_{CS})\chi$ distinguish the
configurations with large $iW_{CS}$ (in QED these are (multi)
photon states with positive helicity). The loop correlation
functions calculated (in a rather formal way) in the states
$\exp(\frac{i}{\hbar}W_{CS})$ are expressed by Gauss linking
numbers \cite{gauss} and Jones polynomials \cite{wittenjones}. We
extend the method of solving the functional Schr\"odinger equation
to non-Abelian gauge fields and to a quantization of massless spin
2 tensor fields. The aim of this method is to approach new
phenomena related to particle's helicity and the linking number in
gauge theories and to try new ways of quantization of gravity.

The plan of the paper is the following. In sec.2 we show that an
addition to the Lagrangian of a total derivative of the term $i\ln
\psi^{g}$, where $\psi^{g}$ is a particular solution of the
Schr\"odinger equation, leads to the solutions in the form
$\psi^{g}\chi$ of the Schr\"odinger equation. In sec.3 we discuss
solutions of the form $\exp(i\frac{W_{CS}}{\hbar})$ in quantum
electrodynamics. We show that general solutions of the
Schr\"odinger equation  are expressed by the solution of a random
self-duality equation. We discuss the corresponding equation for
non-Abelian gauge fields. In sec.4 the method is extended to free
massless spin 2 fields. We briefly discuss a possible extension to
a quantization of gravity.
\section{General framework}

We have discussed a general scheme of a construction of the
solution of the Schr\"odinger equation by means of a diffusion
process in \cite{haba1}\cite{haba2}. In theories with a more
intricate geometry it is interesting to elaborate this
quantization scheme starting from the classical field theory. In
the Lagrangian formulation a change of the Lagrangian $L$ by a
total derivative

\begin{equation} L^{\prime}=L -i\partial_{t}\ln\psi^{g}(A),
\end{equation}
where $\psi^{g}(A)$ is an arbitrary function of the classical
variable $A$, does not change the dynamics. However, the canonical
momentum ( when $L$ is quadratic in $\partial_{t}A$)
 \begin{equation} \Pi^{\prime}=\partial_{t}
A-i\frac{\delta\ln\psi^{g}(A)}{\delta A}
\end{equation}
does change. Then, the canonical Hamiltonian is (changing
$H\rightarrow H^{\prime}$)
\begin{equation}\begin{array}{l}
H^{\prime}=H+i\frac{1}{2}(\Pi^{\prime}\frac{\delta\ln\psi^{g}(A)}{\delta
A}+\Pi^{\prime}\frac{\delta\ln\psi^{g}(A)}{\delta A})
-\frac{1}{2}(\frac{\delta\ln\psi^{g}(A)}{\delta A})^{2}.
\end{array}\end{equation}
The standard quantization sets \begin{equation} \Pi^{\prime}({\bf
x})=-i\hbar\frac{\delta}{\delta A({\bf x})}.
\end{equation}
In eq.(3) we have chosen a particular symmetric  ordering of the
non-commutative terms which leads to the correct quantum
Hamiltonian $H^{\prime}$. As an example, if
\begin{equation}
H=\frac{1}{2}\Pi^{2}+\frac{1}{2}(\nabla A)^{2}+V(A)
\end{equation}
and if $\psi^{g}$ is an eigenstate
\begin{equation}
H\psi^{g}=0
\end{equation}
then from eqs.(3)-(6)
\begin{equation}
H^{\prime}=\int d{\bf x}
\Big(-\frac{\hbar^{2}}{2}\frac{\delta^{2}}{\delta A({\bf x})^{2}}-
\hbar^{2}\frac{\delta\ln\psi^{g}}{\delta A({\bf
x})}\frac{\delta}{\delta A({\bf x})}\Big).
\end{equation}
The Hamiltonian $H^{\prime}$ has the form of a generator of a
diffusion process with a drift determined by $\ln\psi^{g}$. Here,
we have derived the modified Hamiltonian from an addition of the
total derivative to the Lagrangian. Although from the point of
view of the canonical quantization $H^{\prime}$ is as good as $H$
for the canonical quantization we note that $H^{\prime}$ is not
self-adjoint in $L^{2}(dA)$ but in $L^{2}((\psi^{g})^{2}dA)$. In a
quantum formulation we would obtain $H^{\prime}$ by means of the
similarity transformation
\begin{equation}
H^{\prime}=(\psi^{g})^{-1}H\psi^{g}.
\end{equation}  In another way ( which admits time
dependent solutions of the Schr\"odinger equation $\psi_{t}^{g}$)
 we note that if $\psi_{t}^{g}$ is a solution of the Schr\"odinger
 equation $i\hbar\partial_{t}\psi_{t}^{g}=H\psi_{t}^{g}$ and
 $\psi_{t}=\psi_{t}^{g}\chi_{t}$ then $\chi_{t}$ solves the equation
 \begin{equation}
i\hbar\partial_{t}\chi=H^{\prime}\chi.
\end{equation}
The whole scheme assumes that $\ln\psi^{g}$ is a regular function
and the Schr\"odinger equations with $H$ and $H^{\prime}$ have the
unique solutions with the initial values $\psi^{g}$ and
(resp.)$\chi=(\psi^{g})^{-1}\psi$. The correspondence between
$\psi$ and $\chi$ can be broken if $\psi^{g}$ has zeros (we do not
consider such cases in this paper). In quantum mechanics, examples
satisfying eq.(6) are delivered by wave functions of the form
$\psi^{g}=\exp(i\frac{W}{\hbar})$, where $\triangle W=0$ (in
\cite{thooft} such wave functions are distinguished as leading to
the zero cosmological constant) . We could take as $iW$ a real
part of a (pluri)harmonic function. So, in two-dimensions
($z=x_{1}+ix_{2}$)let the potential be
$V=\frac{9}{2}\lambda^{2}(x_{1}^{2}+x_{2}^{2})^{2}$ then
$\psi^{g}=\exp(i\frac{W}{\hbar})$ solves eq.(6) if
$-iW=\frac{\lambda}{2}(z^{3}+\overline{z}^{3})=\lambda(x_{1}^{3}-3x_{1}x_{2}^{2})$.
$\exp(i\frac{W}{\hbar})$ is not square integrable but in eq.(9) we
could look for solution $\chi_{t}$ with the initial condition
$\chi$ decaying sufficiently fast at infinity.

In field theory in one spatial dimension  we could consider
$V=\frac{9}{2}\lambda^{2}(\phi_{1}^{2}+\phi_{2}^{2})^{2}$ and
\begin{equation}-iW=\int dx\Big(\frac{\lambda}{2}((\phi_{1}+i\phi_{2})^{3}+(\phi_{1}-i\phi_{2})^{3})
+\phi_{1}\frac{\partial}{\partial x}\phi_{2}\Big).
\end{equation}
$W$ determines the Hamiltonian as from eq.(6)
\begin{displaymath}
V(\phi)+\frac{1}{2}\partial_{x}\phi_{a}\partial_{x}\phi_{a}=-\frac{1}{2}\frac{\delta
W}{\delta\phi_{a}}\frac{\delta W}{\delta\phi_{a}}
\end{displaymath}
 The last term on the rhs of eq.(10) appears in eq.(1) as a total
derivative of a Lorentz invariant expression (which is also a
topological invariant) in the action $\frac{1}{2}\int
dxdt\epsilon^{\mu\nu}\partial_{\mu}\phi_{1}\partial_{\nu}\phi_{2}$.
Such models are supersymmetric and  have solutions determined by
stochastic differential equations  as discussed by Parisi and
Sourlas \cite{parisisourlas}.

In four space-time dimensions  a similar construction applies to
vector fields ${\bf A}$ which can be considered as
quaternion-valued fields. Let ${\cal A}=\sigma_{k}A_{k}$, where
$\sigma_{k}$ ($k=1,2,3 $) are the Pauli matrices, Then, we
consider a gauge invariant and topological invariant expression
(in the gauge $A_{0}=0$, $\mu=0,1,2,3$)
\begin{displaymath} \frac{1}{4}\int d{\bf
x}\epsilon^{\mu\nu\rho\sigma}\partial_{\mu}A_{\nu}\partial_{\rho}A_{\sigma}
=\partial_{0}\int d{\bf x}Tr(A\partial A)\equiv i\partial_{t}W,
\end{displaymath} where by $\epsilon$ we denote the Levi-Civita antisymmetric
symbols in various dimensions and the quaternionic derivative is
$\partial=\sigma_{k}\partial_{k}$. We obtain the Chern-Simons wave
function $\exp(i\frac{W}{\hbar})$ which is a (non-normalizable)
ground state solution (6) for the quantum electromagnetic field.
This model will be discussed in detail in the next section.
\section{Hamiltonian
evolution of the electromagnetic field}

We consider now the Schr\"odinger equation for the electromagnetic
field $A_{\mu}$ with the Lagrangian
($F_{\mu\nu}=\partial_{\mu}A_{\nu}-\partial_{\nu}A_{\mu}$)
\begin{displaymath}
L=-\frac{1}{4}F_{\mu\nu}F^{\mu\nu}
\end{displaymath}
We quantize the model in the radiation gauge ($A_{0}=0$ and
$\partial _{j}A_{j}=0$). The Hamiltonian is
\begin{equation}
H=\frac{1}{2}\sum_{j}\Pi_{j}^{2}+\frac{1}{2}(\nabla\times {\bf
A})^{2}.
\end{equation}
We expand  $A_{j}$ in amplitudes of polarization $\lambda$
(\cite{mandel}, sec.10.2.2, see also \cite{gross})
\begin{equation}\begin{array}{l} A_{j}=(2\pi)^{-\frac{3}{2}}\sum_{\lambda}\int \frac{d{\bf k}}{2\vert{\bf k}\vert}\Big(u_{\lambda}({\bf
k})e_{j}({\bf k},\lambda)\exp(i{\bf kx}-i\omega
t)\cr+(u_{\lambda}({\bf k}))^{*}e_{j}^{*}({\bf
k},\lambda)\exp(-i{\bf kx}+i\omega
t)\Big)=\sum_{\lambda}A_{j}^{(\lambda)}+(A_{j}^{(\lambda)})^{*},
\end{array}\end{equation} where $k=\vert{\bf
k}\vert=\omega$ and we apply a decomposition of the field into the
right and left (complex) polarization vectors which can be
expressed by real plane polarization vectors ${\bf e}^{(x)}$ and
${\bf e}^{(y)}$
\begin{displaymath}
e_{j}({\bf k},+)=\frac{1}{\sqrt{2}}(e_{j}^{(x)}({\bf
k})+ie_{j}^{(y)}({\bf k})),
\end{displaymath}\begin{displaymath}
e_{j}({\bf k},-)=\frac{1}{\sqrt{2}}(e_{j}^{(y)}({\bf
k})-ie_{j}^{(x)}({\bf k})).
\end{displaymath}
$e_{j}^{(x)}$ and $e_{j}^{(y)}$ describe the plane polarizations
of the electromagnetic wave moving in the direction ${\bf k}$(
i.e.,${\bf k}{\bf e}=0$). We have ${\bf k}\times {\bf
e}^{(x)}=k{\bf e}^{(y)}$ and ${\bf k}\times {\bf e}^{(y)}=-k{\bf
e}^{(x)}$. Hence, ${\bf k}\times ({\bf e}^{(x)}+i{\bf
e}^{(y)})=-ik({\bf e}^{(x)}+i{\bf e}^{(y)}$)  . As a consequence,
the components with a definite polarization satisfy the equation
($\lambda=\pm 1$,$\epsilon(\lambda)$ is the sign of $\lambda$)
\begin{equation}
\partial_{t}{\bf A}^{(\lambda)}=i\epsilon(\lambda)\nabla\times {\bf A}^{(\lambda)}
\end{equation}
which coincides  with the (anti) self-duality condition depending
on the sign $\epsilon(\lambda)$ of the circular polarization
(helicity). So, $A_{j}$ is a sum of self-dual and anti self-dual
pieces. $u_{\lambda}$ ($u_{\lambda}^{*}$) become the annihilation
(creation) operators in the canonical quantization. When we
calculate the correlation function in the Fock vacuum $\Omega$ of
the pieces of the electromagnetic field with a definite
polarization then we obtain
\begin{equation}\begin{array}{l}
S_{jl}(+)=(\Omega,A_{j}^{(+)}({\bf k},t)(A_{l}^{(+)}({\bf
k}^{\prime}))^{+}\Omega)\cr= (2\pi)^{-3}\delta({\bf k}+{\bf
k}^{\prime})(2k)^{-1}\exp(-i\vert{\bf k}\vert t) e_{j}({\bf
k},+)(e_{l}({\bf k},+))^{*}.\end{array}
\end{equation}
It follows that
\begin{equation}
\partial_{t}S_{jl}(+)=-\epsilon _{jmn}k^{m}S_{nl}(+).
\end{equation}
The correlation function (14) of the polarized pieces of the
electromagnetic field depends on the choice of the polarization
vectors. The solution $S_{nl}$ of the self-duality equation (15)
is not unique as well. It depends on the initial condition and  on
an arbitrary longitudinal tensor proportional to $k_{n}k_{l}$. It
is well-known that the self-duality is related to the photon
helicity \cite{bialynicki}\cite{ashtekar2}. Summing over the
polarizations we obtain the projection operator onto the
transverse modes
\begin{equation}
\sum_{\lambda}e_{j}({\bf k},\lambda)(e_{l}({\bf k},\lambda))^{*}=
\delta_{jl}-k_{j}k_{l}k^{-2}\equiv P_{jl}.\end{equation}The
two-point correlation function in the Fock vacuum is
\begin{equation}
(\Omega,A_{j}({\bf k},t)A_{l}({\bf
k}^{\prime})\Omega)=\sum_{\lambda}S_{jl}(\lambda) =
(2\pi)^{-3}\delta({\bf k}+{\bf
k}^{\prime})(2k)^{-1}\exp(-i\vert{\bf k}\vert t)P_{jl}({\bf k})
\end{equation}
 The self-dual
expansion has been generalized  by Weinberg
\cite{weinberg1}\cite{weinberg2} to massless bosons with higher
spin.

 The quantum version of the Hamiltonian $H$ (11) in the radiation gauge
 results from  the classical Lagrangian (the projector
 $P_{jk}$ (16) signals that only transverse components serve as
 dynamical variables; for a theory of functional diffusion-type differential equations see \cite{daletsky}\cite{elworthy})
\begin{equation} H=\int d{\bf x}
\Big(-\frac{\hbar^{2}}{2}\frac{\delta}{\delta A_{j}({\bf
x})}P_{jk}\frac{\delta}{\delta A_{k}({\bf
x})}+\frac{1}{2}(\nabla\times {\bf A})^{2}\Big).
\end{equation}Now, we can apply the framework expressed by eqs.(8)-(9). Let $\psi^{g}$ be a particular time independent solution  of the
Schr\"odinger equation
\begin{equation}
i\hbar\partial_{t}\psi=H\psi
\end{equation}
We can write a general solution in the form \begin{equation}
\psi_{t}=\psi^{g}\chi_{t},
\end{equation}
where $\chi_{t}$ is the solution of the equation (cp. with eq.(7))
\begin{equation}
\partial_{t}\chi_{t}=i\hbar\int d{\bf x} \Big(\frac{1}{2}\frac{\delta}{\delta A_{j}({\bf
x})}P_{jk}\frac{\delta}{\delta A_{k}({\bf x})}
+\frac{\delta\ln\psi^{g}}{\delta A_{j}({\bf
x})}P_{jk}\frac{\delta}{\delta A_{k}({\bf x})}\Big)\chi_{t}.
\end{equation}
The solution of eq.(21) can be expressed as
\begin{equation}
\chi_{t}({\bf A})=E[\chi({\bf A}_{t}({\bf A}))],
\end{equation}
where $\chi$ is the initial value for eq.(21) and ${\bf
A}_{t}({\bf A})$ is the solution (with the initial condition ${\bf
A}$) of the stochastic equation
\begin{equation}
d{\bf A}=i\hbar \frac{\delta}{\delta {\bf A}({\bf
x})}\ln\psi^{g}dt+\sqrt{i\hbar}d{\bf B},\end{equation} where the
Brownian motion ${\bf B}(t,{\bf x})$ is the Gaussian process with
the covariance
\begin{equation}
E[B_{k}(t,{\bf x}),B_{l}(s,{\bf
y})]=(\delta_{kl}-\partial_{k}\partial_{l}\triangle^{-1})\delta({\bf
x}-{\bf y})min(t,s).
\end{equation}

 We distinguish two particular
solutions of the Schr\"odinger equation (19): the ground state
solution (with an infinite ground state energy)
\begin{equation}
\psi_{0}^{g}=\exp\Big(-\frac{1}{2\hbar}\int {\bf
A}(-\triangle)^{\frac{1}{2}}{\bf A}\Big)\equiv
\exp(-\frac{W_{0}}{\hbar})
\end{equation}
and the Chern-Simons solution (with zero energy)
\begin{equation}
\psi_{CS}=\exp(-\frac{1}{2\hbar}\int d{\bf
x}A_{j}\epsilon^{jkl}\partial_{k}A_{l})\equiv
\exp(-\frac{W_{CS}}{\hbar}).
\end{equation}

The stochastic equation (23) for the ground state solution is

\begin{equation}
d{\bf A}=-i(-\triangle)^{\frac{1}{2}} {\bf A}({\bf
x})dt+\sqrt{i\hbar}d{\bf B},
\end{equation}

The solution of eq.(27) is
\begin{equation}
{\bf A}_{t}=\exp(-it\sqrt{-\triangle}){\bf
A}+\sqrt{i\hbar}\int_{0}^{t}\exp(-i(t-s)\sqrt{-\triangle})dB_{s}.
\end{equation}
We obtain the standard quantum field theory of the quantum
(transverse) electromagnetic field ${\bf A}^{Q}$ (which is defined
by the time-ordered two-point correlation function) calculating
the time ordered correlation function (17) in the Fock vacuum on
the lhs and the stochastic expectation value on the rhs (using
eq.(28))
\begin{equation}\begin{array}{l}
(\psi_{0}^{g}, T(A_{j}^{Q}(t,{\bf x})A_{k}^{Q}(s,{\bf
y}))\psi_{0}^{g})=\int d{\bf
A}\vert\psi_{0}^{g}\vert^{2}E[A_{j}(t,{\bf y})A_{k}(s,{\bf x})]
\cr=\frac{1}{2}P_{jk}\Big((-\triangle)^{-\frac{1}{2}} \exp(-i\vert
t-s\vert (-\triangle)^{\frac{1}{2}})\Big)({\bf x},{\bf y}).
\end{array}
\end{equation}
The stochastic equation (23) for the Chern-Simons solution (26)
can be expressed in the form
\begin{equation}
dA_{j}=i\epsilon_{jkl}\partial_{k}A_{l}dt +\sqrt{i\hbar}dB_{j}.
\end{equation}
Eq.(30) can be considered as a random perturbation $w$ of the
self-duality condition
\begin{equation}
F_{\mu\nu}=\frac{i}{2}\epsilon_{\mu\nu\alpha\beta}F^{\alpha\beta}dt
+\sqrt{i\hbar}w_{\mu\nu},
\end{equation}where
$F_{\mu\nu}=\partial_{\mu}A_{\nu}-\partial_{\nu}A_{\mu}$. In fact,
we obtain eq.(30) setting $\mu=0$, $\nu=k$ and
$w^{0k}=\frac{dB^{k}}{dt}$ in eq.(31).

 In order to find a solution of eq.(30)  let us transform it to the momentum space. Then, consider the
antisymmetric real matrix
\begin{displaymath} M_{jk}=\epsilon_{jkl}p_{l}.
\end{displaymath}
 $O(s)\equiv \exp(-sM)$ is an orthogonal matrix with the matrix
elements
\begin{equation}
O(s)_{jk}=(\delta_{jk}-p_{j}p_{k} {\bf p}^{-2})\cos(s\vert{\bf
p}\vert)-\epsilon_{jkl}p_{l}\vert {\bf p}\vert^{-1}\sin(s\vert
{\bf p}\vert).
\end{equation}
The correlation functions of quantum fields $A_{t}^{Q}$ are
determined by the stochastic process as (for multitime correlation
functions see \cite{haba2}) \begin{equation}\begin{array}{l}
(\psi^{g}\chi,{\bf A}_{t}^{Q}({\bf x}){\bf A}^{Q}({\bf
y})\psi^{g}\chi)=
 (\psi_{t},{\bf A}({\bf x})
E[{\bf A}_{t}({\bf y})\chi({\bf A}_{t})]\psi^{g})
\end{array}\end{equation}where $\psi_{t}=U_{t}(\psi^{g}\chi)=\psi^{g}\chi_{t}$
and $U_{t}$ is the unitary evolution determined by the
Schr\"odinger equation (19). For the Chern-Simons wave function
(26) on a formal level (which can be justified by using a
regularization) when $\chi\rightarrow 1$
\begin{equation}\begin{array}{l}
 (\psi_{t},{\bf A}({\bf x})
E[{\bf A}_{t}({\bf y})\chi({\bf A}_{t})]\psi_{CS})\rightarrow
(\psi_{CS},{\bf A}({\bf x}) E[{\bf A}_{t}({\bf
y})]\psi_{CS})\cr\simeq (2M)^{-1}\exp(-tM )({\bf x},{\bf
y}).\end{array}
\end{equation}The lhs of eq.(34) is
well-defined when $\chi({\bf A})$ is decaying fast for  large
${\bf A}$. We discuss the limit (34) in the Appendix for a special
class of Gaussian regularizing functions $\chi$. We note that
$M^{-1}_{jk}=\epsilon_{jkl}\partial_{l}\triangle^{-1}$ in the
correlation function of the electromagnetic potentials  leads to
the Gauss linking number in the correlation functions of the
electromagnetic potential integrated over closed loops
\cite{gauss}\cite{wittenjones}.

 The solution of
eq.(30) for the  field ${\bf A}_{t}$ in the momentum space with an
initial condition ${\bf A}$ has the form
\begin{equation}
{\bf A}(t)=O(t){\bf A}+\sqrt{i\hbar}\int_{0}^{t}O(t-\tau)d{\bf
B}(\tau).
\end{equation}
The first term $O(t){\bf A}$ in eq.(35) is the solution of the
self-duality condition ( this is also a complex solution of the
Maxwell equations).It appears in the limit $\hbar\rightarrow 0$ of
the solution of the Schr\"odinger equation (22).

Using the solution (35) we can obtain  solutions of the
Schr\"odinger equation describing the states with distinguished
positive helicity ( when $\chi$ has the same probability
distribution for both helicities). For states with a negative
helicity we would use the Chern-Simons states with an opposite
sign  of $W$ and anti-self-dual stochastic equations.

  For  a calculation of expectation values (33) we need the
  covariance of  the stochastic field ${\bf A}_{t}$
\begin{equation}\begin{array}{l}
E[A_{j}(t,{\bf p})A_{k}(s,{\bf p}^{\prime})]=(O(t){\bf A}({\bf
p}))_{j}(O(s){\bf A}({\bf p}))_{k} \cr
+i\hbar\int_{0}^{min(t,s)}O(t-\tau)_{jr}O(s-\tau)_{rk}d\tau\cr
=O_{jr}(t)O_{kn}(s)A_{r}({\bf p})A_{n}({\bf p}^{\prime})+i\hbar
G_{jk}(t,s)\delta({\bf p}+{\bf p}^{\prime})
\end{array}
\end{equation}
where
\begin{equation}
G_{jk}(t,s)=\int_{0}^{ min(t,s)}O_{jk}(t+s-2\tau)d\tau
\end{equation}
Evaluation of the integral (37)  on transverse states ( then the
longitudinal arbitrariness is irrelevant) gives
\begin{equation}\begin{array}{l} G_{jk}(t,s)=\frac{1}{2}\Big(M^{-1}(\exp (-\vert t-s\vert
M)-\exp (-M(t+s))\Big)_{jk}\cr=\frac{1}{2}\Big(M^{-1}(O (\vert
t-s\vert )-O(t+s))\Big)_{jk}.\end{array}
\end{equation}
For the ground state process (25) we would have $M=\vert{\bf
p}\vert$ and  $O(t)=\exp(-it\vert {\bf p}\vert)$.

The stochastic formalism can be generalized to non-Abelian gauge
theories.  We consider in Minkowski space-time an algebra-valued
vector field $A_{\mu}=\sum A_{\mu}^{a}\tau^{a}$ where $\tau^{a}$
form a basis of the Lie algebra of a compact Lie group. The
self-duality equation in non-Abelian gauge theory has the same
form as in the Abelian case
\begin{equation}
F^{a\mu\nu}=\frac{i}{2}\epsilon^{\mu\nu\sigma\rho}F^{a}_{\sigma\rho}
\end{equation}
but now
\begin{equation}
F_{\mu\nu}^{a}=\partial_{\mu}A_{\nu}^{a}-\partial_{\nu}A_{\mu}^{a}+g
f_{abc}A^{b}_{\mu}A^{c}_{\nu},
\end{equation} where $g$ is a coupling constant and $f_{abc}$ are the (real) structure constants
of the Lie algebra. We add a noise $w$ to eq.(39)
\begin{equation}
F^{a\mu\nu}=\frac{i}{2}\epsilon_{\mu\nu\sigma\rho}F^{a}_{\sigma\rho}+\sqrt{i\hbar}w^{a}_{\mu\nu}
\end{equation}
The covariant derivative of both sides of eq.(41) is
\begin{equation}
\nabla^{\mu}F^{a}_{\mu\nu}=\frac{i}{2}\epsilon^{\mu\nu\sigma\rho}\nabla^{\mu}F^{a}_{\sigma\rho}
+\sqrt{i\hbar}\nabla^{\mu}w^{a}_{\mu\nu},
\end{equation}
where
\begin{equation}
\nabla^{\mu}F^{a}_{\sigma\rho}=\partial^{\mu}F^{a}_{\sigma\rho}
+gf_{abc}A^{b\mu }F^{c}_{\sigma\rho}.
\end{equation}
The first term on the rhs of eq.(42) is vanishing on the basis of
the Jacobi identity. Hence, without the noise equations of motion
for the Yang-Mills field are satisfied. Then, with the noise
\begin{displaymath}
\nabla^{\mu}F^{a}_{\mu\nu}=\sqrt{i\hbar}\nabla^{\mu}w^{a}_{\mu\nu}.
\end{displaymath}
Choosing the noise in such a way that
$\nabla^{\mu}w^{a}_{\mu\nu}=0$
 we ensure the fulfillment of the Gauss constraint $\nabla^{k}F_{k0}=0$.

 We can explain the change of the noise $w$ on the basis of the theory of the stochastic equations on a manifold
 \cite{elworthy}\cite{williams}\cite{lewis}.  We have
redundant degrees of freedom in gauge theory. We choose the
temporal gauge for quantization. The self-duality equation is
still invariant under an infinite dimensional group of
space-dependent gauge transformations. For quantization we should
consider a Markov process on the coset space ${\cal A}/{\cal G}$
of the linear space of connections ${\cal A}$ divided by the group
of gauge transformations ${\cal G}$. If the stochastic process is
to run over  ${\cal A}/{\cal G}$ then we should project $w$ onto
the tangent space of ${\cal A}/{\cal G}$. The projector is
\cite{babelon}
\begin{equation}
P_{jk}=\delta_{jk}-\nabla_{j}{\cal D}\nabla_{k},
\end{equation}
where
\begin{equation}
{\cal D}=(\nabla_{j}\nabla^{j})^{-1}.
\end{equation}
In such a case eq.(41) reads
\begin{equation}
dA^{a}_{j}=\frac{i}{2}\epsilon_{jkl}F^{a}_{kl}dt+\sqrt{i\hbar}P_{jk}dB^{a}_{k}
\end{equation} where ${\bf B}^{a}$ are independent Brownian
motions with the covariance
\begin{displaymath}
E[B_{j}^{a}(t,{\bf x})B_{k}^{b}(s,{\bf
y})]=\delta^{ab}\delta_{jk}min(t,s)\delta({\bf x}-{\bf y})
\end{displaymath}
It follows that the Gauss constraint
\begin{equation}
\nabla^{j}F^{a}_{j0}=\nabla^{j}\partial_{t}A^{a}_{j}=0
\end{equation}
is satisfied as a consequence of eq.(46).

The Schr\"odinger  equation following from eq.(46)
is\begin{equation}
\partial_{t}\chi=\int d{\bf x} \Big(\frac{i\hbar}{2}\frac{\delta}{\delta
{\bf A}({\bf x})}\int d{\bf y}P({\bf x},{\bf
y})\frac{\delta}{\delta {\bf A}({\bf
y})}+\frac{i}{2}\epsilon_{jkl}F_{kl}\frac{\delta}{\delta A_{l}(
{\bf x})}\Big)\chi.
\end{equation}
If\begin{equation} \psi_{CS}=\exp(\frac{1}{2\hbar}Tr\int ({\bf
A}\wedge d{\bf A}+\frac{2}{3}{\bf A}\wedge {\bf A}\wedge {\bf
A})\Big)
\end{equation}then
\begin{equation}
\psi_{t}=\psi_{CS}\chi_{t}
\end{equation}
solves the Schr\"odinger equation
\begin{equation}
i\hbar\partial_{t}\psi=\int d{\bf x}
\Big(-\frac{\hbar^{2}}{2}\frac{\delta}{\delta {\bf A}({\bf
x})}\int d{\bf y}P({\bf x},{\bf y})\frac{\delta}{\delta {\bf
A}({\bf y})}+\frac{1}{4}F_{kl}^{a}F_{kl}^{a}\Big)\psi.
\end{equation}because
$\psi_{CS}$ is a time-independent solution of eq.(51). The
functional Schr\"odinger equation (51) is discussed in
\cite{babelon}\cite{mitter2}\cite{gawedzki}. It needs a
regularization because of the ultraviolet problems of the quantum
Yang-Mills theory.

The solution of  eq.(48) is
\begin{equation}
\chi_{t}({\bf A})=E[\chi({\bf A}_{t}({\bf A}))]
\end{equation}
where ${\bf A}_{t}({\bf A})$ is the solution of the stochastic
equation (46).

By means of the Cameron-Martin formula
\cite{elworthy}\cite{williams} we may return (on a formal level)
to the standard form of the Feynman functional integral for the
Yang-Mills theory with the Lagrangian
$-\frac{1}{4}F^{a\mu\nu}F_{a\mu\nu}$. In the intermediate step we
use the fact that
$\frac{1}{2}F^{*}F=\partial_{\mu}(\epsilon^{\mu\nu\sigma\rho}A_{\nu}F_{\sigma\rho})$
in order to extract the CS state in eq.(50) as discussed in sec.2.

We can solve the stochastic equation by means of a perturbation
theory in the coupling $g$. So, at $g=0$ we get  Abelian vector
fields. The virtue of the stochastic formulation (46) and (52) may
be revealed in numerical calculations when there are standard
numerical methods for solutions of stochastic differential
equations. If  the spatial integral in eq.(49) is over a finite
region or over a manifold with a boundary then $\psi_{CS}$  is not
invariant under gauge transformations which do not vanish on the
boundary. In such a case these gauge degrees of freedom on the
boundary must be quantized as well (a Wess-Zumino-Witten type
model, see \cite{knot})
\section{Quantization of massless spin 2 tensor fields}
We could quantize gravitational waves (following the approach of
secs.2-3) through a construction of the Hamiltonian for linearized
Einstein gravity. Another way is to follow Weinberg's quantization
of spin 2  massless fields \cite{weinberg1}\cite{weinberg2}. We
approach the problem from the non-perturbative ADM formulation
\cite{adm} of the canonical gravity which leads to the
Wheeler-deWitt (WdW) equation \cite{dewitt}. We hope that such an
approach can lead to a quantum theory beyond the linear
approximation. The WdW equation is a consequence of the invariance
of general relativity under the diffeomorphism transformations. In
quantum gravity the Hamiltonian $H$ of a total system (matter
+gravity) acting on arbitrary states $\psi$ gives zero (expressed
in the units $l^{2}_{p}=16\pi G=c=\hbar=1$, where $G$ is the
gravitational constant and $l_{p}$ is the Planck length)
\begin{equation}
H\psi=\Big(\int d{\bf x} \frac{\delta}{\delta \gamma_{jl}({\bf
x})}G_{jl;kn}\frac{\delta}{\delta \gamma_{kn}({\bf x})}+ \int
d{\bf x}\sqrt{\gamma}R-H_{m}\Big)\psi=0,
\end{equation}

\begin{equation}G_{ij;mn}
=\frac{1}{2}\gamma^{-\frac{1}{2}}(\gamma_{im}\gamma_{jn}+\gamma_{in}\gamma_{jm}
-\gamma_{ij}\gamma_{nm}).
\end{equation}
$\gamma=\det(\gamma_{jk})$, $\gamma_{jk}$ is the metric tensor on
the time zero surface of the Riemannian manifold, $R$ is the
scalar curvature on this surface. $H_{m}$ is the matter
Hamiltonian, so that $H_{m}\psi$ could possibly be replaced by a
time derivative $i\partial_{t}\psi$ in a semiclassical
approximation of the matter  fields \cite{kiefer}.$G_{ij;mn}$ is
the (ultralocal) metric on the space of Riemannian metrics
$\gamma_{jk}$(the "superspace"
)\cite{friedman}\cite{hart2}\cite{giulini}. We look for solutions
of the WdW equation with $H_{m}=0$ in a linear approximation to
gravity. In a more general context (beyond the linear
approximation), we point out that finding a particular solution of
the WdW equation (53) with $H_{m}=0$ could enable a derivation of
all other solutions with $H_{m}=0$ and subsequently a solution
with $H_{m}\psi$ replaced by $i\partial_{t}\psi$. The method is to
replace the equation $H\psi=0$ by $i\partial_{t}\psi=H\psi$ and
look for the limit $t\rightarrow \infty$ of the time-dependent
Schr\"odinger equation.

The curvature term in eq.(53) in the linear approximation for the
metric
\begin{equation}
\gamma_{jk}=\delta_{jk}+h_{jk} \end{equation}
 up to quadratic terms in $h$ is
\begin{equation}
R=\gamma^{jk}R_{jk}=\frac{1}{4}(
\delta^{jk}+h^{jk})(h^{r}_{j,rk}+h^{r}_{k,rj}-h^{,r}_{jk,r}
-h^{l}_{l,rr}).
\end{equation}
The  term linear in $h$ in eq.(53) coming from the $\int d{\bf
x}\sqrt{\gamma}R$ term is

\begin{equation}
\partial_{j}\partial_{k}h^{jk}-\triangle h^{j}_{j}.
\end{equation}We  look for $\psi=\exp(-W)$
such that the second derivative in eq.(53) is quadratic in $h$.
Then, the linear term should be equal to zero. The
transverse-traceless (TT) gauge
\begin{equation}
\partial_{j}h^{jk}=0, h^{j}_{j}=0
\end{equation}
ensures the vanishing of the term (57) in the linear approximation
(55)(see the discussion in \cite{kuchar}\cite{hartle} that the
TT-metric components  supply the coordinates on the manifold of
metrics on the time-zero surface of the pseudo-Riemannian
manifold).

The second order term in $\int d{\bf x}\sqrt{\gamma}R$ in the
transverse-traceless gauge is
\begin{equation}
\int d{\bf x}\sqrt{\gamma}R=-\frac{1}{4}\int d{\bf
x}h^{TT}_{jk}\triangle h^{TT}_{jk}=\frac{1}{4}\int d{\bf x}\nabla
h^{TT}_{jk}\nabla h^{TT}_{jk}.
\end{equation}
After a choice of coordinates on the superspace we must express
the metric $G$ in these new coordinates. This is similar to the
modification of eqs.(21),(44) and (51) in gauge theories. We
rewrite eq.(53) expressing the transverse-traceless tensor
$h^{TT}_{jk} $ as a projection $\Lambda$ of an arbitrary symmetric
tensor $h_{mn}$
\begin{equation}
\gamma_{jk}=\delta_{jk}+\Lambda_{jk;mn}h_{mn},
\end{equation}
where
\begin{equation}\Lambda_{ij;mn}
=\frac{1}{2}(P_{im}P_{jn}+P_{in}P_{jm} -P_{ij}P_{nm})
\end{equation}with
\begin{displaymath}
P_{jk}=\delta_{jk}-\partial_{j}\partial_{k}\triangle^{-1}.
\end{displaymath}
 Now,
\begin{equation}
\frac{\delta}{\delta h_{jk}}=\frac{\delta \gamma_{mn}}{\delta
h_{jk}} \frac{\delta}{\delta \gamma_{mn}}.
\end{equation}
Working in the lowest order approximation in $h$ in eq.(53) we
approximate $G$ (54) by
\begin{equation}G_{ij;mn}
=\frac{1}{2}(\delta_{im}\delta_{jn}+\delta_{in}\delta_{jm}
-\delta_{ij}\delta_{nm}).
\end{equation} After the change to the TT coordinates (58)
using the relation
\begin{displaymath}
G_{jk;mn}\Lambda_{mn;rl}=\Lambda_{jk;rl}.
\end{displaymath}
 we obtain the Hamiltonian

\begin{equation} H=\int d{\bf x}
\Big(-\frac{\delta}{\delta h_{jl}({\bf x})}\int d{\bf
y}\Lambda_{jl;kn}({\bf x},{\bf y})\frac{\delta}{\delta h_{kn}({\bf
y})}+\frac{1}{4}(\nabla h_{jl})^{2}\Big).
\end{equation}
The (normalizable) ground state of this Hamiltonian (with a
subtracted infinite ground state energy) is
\begin{equation}
\psi^{g}=\exp\Big(-\frac{1}{4}\int d{\bf
x}h_{jl}^{TT}\sqrt{-\triangle}h_{jl}^{TT}\Big).
\end{equation}
There is also the Chern-Simons ground state satisfying the WdW
equation which is not normalizable. It has the form
\begin{equation}
\psi_{CS}=\exp\Big(-\frac{1}{4}\int d{\bf
x}h_{jl}^{TT}\epsilon_{jmn}\partial_{m}h_{nl}^{TT}\Big).
\end{equation}
Owing to the $TT$-condition (58) this state is invariant under the
infinitesimal change of coordinates ( if $h_{jl}$ and $\xi_{j}$
vanish at infinity)
\begin{displaymath}
h_{jl}^{TT}\rightarrow h_{jl}^{TT}+\xi_{j,l}+\xi_{l,j}
\end{displaymath}
In this state the correlation function $<hh>$ gives the linking
number for integrals over closed curves \cite{nicolai} (as in the
case of gauge fields the inverse of the CS operator gives the
Gauss linking number).

The stochastic quantization scheme (23) leads to the stochastic
equation
\begin{equation}
dh_{jk}=-i\sqrt{-\triangle}h_{jk}dt+\sqrt{i}dB_{jk}
\end{equation}
for the ground state (65) whereas for the CS state (66) we obtain
\begin{equation}
dh_{jl}=i\epsilon_{jmn}\partial_{m}h_{nl}+\sqrt{i}dB_{jl},
\end{equation}
where $B_{jl}$ is the Brownian motion with the correlation
function
\begin{equation}
E[B_{ij}(t,{\bf x})B_{mn}(s,{\bf
y})]=2\Lambda_{ij;mn}min(t,s)\delta({\bf x}-{\bf y}).
\end{equation} It
follows from eqs. (67)-(69) that if we choose a
transverse-traceless tensor as an initial condition then at any
time the tensor will remain transverse-traceless.

The solution of eqs.(67)-(68) is
\begin{displaymath}
h_{t}=O(t)h+\sqrt{i\hbar}\int_{0}^{t}O(t-s)dB_{s},
\end{displaymath} where $O(t)=\exp(-tM)$ with $M=\vert{\bf p}\vert$
for the ground state (65) and $M_{jk}=\epsilon_{jkr}p_{r}$ for the
$CS$ state (66).
 Then,
the solution of the evolution equation $i\partial_{t}\psi=H\psi$
where $\psi_{t}=\psi_{CS}\chi_{t}$ with the initial condition
$\psi_{CS}\chi$ is
\begin{equation}
\chi_{t}=E[\chi(h_{t}(h))].
\end{equation}
We could use this equation to derive other solutions of the WdW
equation (53) $H(\psi_{CS}\chi)=0$ as a limit $t\rightarrow \infty
$ of $\psi_{CS}\chi_{t}$ (this is similar to the Parisi-Wu
quantization \cite{parisiwu}).

It is known
\cite{polar0}\cite{polar1}\cite{polar2}\cite{bialynicki}\cite{polar00}
that the gravitational waves of definite helicity in the
transverse-traceless gauge satisfy the self-duality equation. In
detail,  we decompose the real transverse-traceless  metric in
complex components with definite helicity as
\begin{equation}\begin{array}{l}
h_{jk}({\bf x},t)=h_{jk}^{(+)}({\bf x},t)+h_{jk}^{(-)}({\bf
x},t)\cr =(2\pi)^{-\frac{3}{2}}\int d{\bf p}(2\vert{\bf
p}\vert)^{-1}(e_{jk}(t,{\bf p})\exp(i{\bf
px})+\overline{e_{jk}(t,{\bf p})} \exp(-i{\bf
px})),\end{array}\end{equation} where
\begin{equation}
e_{jk}(t,{\bf p})=e_{jk}^{+}(t,{\bf p})+ie_{jk}^{\times}(t,{\bf
p})
\end{equation} describes the circular polarization.
Relating the representation (72) to the case of photon
polarization (12) we may write
\begin{displaymath}
e_{jk}^{(+)}=\frac{1}{2}(e^{(x)}_{j}e^{(y)}_{k}+e^{(y)}_{j}e^{(x)}_{k}),
\end{displaymath}
\begin{displaymath}
e_{jk}^{(\times)}=\frac{1}{2}(e^{(x)}_{j}e^{(y)}_{k}-e^{(y)}_{j}e^{(x)}_{k}).
\end{displaymath}
 Then
\begin{equation}
\partial_{t}e_{jk}^{(+)}(t,{\bf p})=\epsilon_{jln}p_{n}e_{lk}^{(+)}(t,{\bf p})
\end{equation}
Hence
\begin{equation}
\partial_{t}h_{kl}^{(+)}=i\epsilon_{kjn}\partial_{n}h_{jl}^{(+)}
\end{equation}for positive helicity and \begin{equation}
\partial_{t}h_{kl}^{(-)}=-i\epsilon_{kjn}\partial_{n}h_{jl}^{(-)}
\end{equation}
for negative  helicity. Hence, the quantization based on the
Hamiltonian (64) and the CS ground state (66) leads to a
quantization of gravitons with a definite helicity (similar to the
case of photons of sec.3, so that the analogs of eqs.(14),(29)
(33), (34) remain valid for graviton states). The Gaussian noise
resulting from the creation-annihilation operators is replaced by
the Brownian motion.

We could write the stochastic self-duality equation (68) in terms
of the tetrads $e^{a}_{j}$ satisfying the relation
\begin{equation}
\gamma_{jk}=e^{a}_{j}e^{a}_{k}.
\end{equation}
In the linear approximation (55)
($e^{a}_{k}=\delta^{a}_{k}+\frac{1}{2}h_{ka}$ ) the corresponding
self-duality equation is the same as for the gauge fields
(eqs.(30) and (46)).It is shown in detail in \cite{krasnov} that
in the linear approximation in the $TT$-coordinates the gravity is
equivalent to the $SO(3)$ gauge theory ( the index $a$ refers to
the so(3) algebra).
 For general $\gamma_{jk}$ we cannot express the Chern-Simons term $W_{CS}$ in
the wave function (66) by means of the metric in a way invariant
under diffeomorphisms. When we express $\gamma_{jk}$ in eq. (76)
in terms of the tetrads $e^{a}_{j}$ then in the linear
approximation there are various candidates for $W_{CS}$. The
simplest  is
\begin{equation}
W_{CS}=\int d{\bf x}\epsilon^{ijk}e_{ia}\partial_{k}e_{ja}
\end{equation} It can also be expressed  as
\begin{equation}
W_{CS}=\int d{\bf
x}\epsilon^{ijk}e_{ia}e_{la}e_{jb}\partial_{k}e^{lb}
\end{equation}
This is exactly the formula (2.8) of Kodama \cite{kodama} ( see
also \cite{kodama2}) who is using $W$ in his suggestion for the
scalar product in the Hilbert space of wave functions in quantum
gravity in the Ashtekar formulation \cite{ashtekar}
\begin{equation}
(\chi_{1},\chi_{2})=\int
de_{j}^{a}\exp(-\frac{2W_{CS}}{\hbar})\overline{\chi_{1}}\chi_{2}
\end{equation}
The scalar product (79) is applied in the definition of the
Hermitian part of Ashtekar's connection . The expression (77) is
invariant under a change of coordinates but is not invariant under
local rotations ${\cal O}$. We need an $O(3)$ connection $\omega$
in order to make (77) covariant (in order to reduce the number of
degrees of freedom of $e^{a}_{j}$ from 9 to 6). Now,
\begin{equation}
W^{\omega}_{CS}\equiv \int d{\bf x}e\wedge d_{\omega}e
+S(\omega)=\int d{\bf
x}\epsilon^{ijk}e_{ia}(\delta^{ab}\partial_{k}+\omega_{k}^{ab})e_{jb}+S(\omega)
\end{equation}
is invariant under the local transformation ${\cal O}\in O(3)$
\begin{displaymath} e\rightarrow {\cal O}e
\end{displaymath}
\begin{displaymath}
\omega\rightarrow {\cal O}\omega {\cal O}^{-1}-\partial {\cal
O}{\cal O}^{-1}
\end{displaymath}
if the action $S(\omega)$  is invariant. We can construct the
connection $\omega_{j}$ from the requirement of zero torsion (as
in Schwinger's approach to gravity \cite{schwinger}).
Then,$\omega_{j}$ is expressed as a linear function of the
derivatives of $e^{a}_{j}$. We could define the Hamiltonian using
the wave function $\psi_{CS}=\exp(-W_{CS})$ and the requirement
$H\psi_{CS}=0$ in eq.(53)(when the second derivative term in  $H$
(53) is defined then the remaining part of the Hamiltonian is
determined by $H\psi_{CS}=0$ as pointed out in sec.2) . This
Hamiltonian would depend on the choice of $S(\omega)$. It seems
unlikely that a proper choice of $S$ would give the Einstein
Hamiltonian in tetrad formulation (see\cite{castel}). We expect
that it may lead to some models of teleparallel gravity as
reviewed in \cite{maluf}. These problems are now under
investigation.

\section{Summary}
We have discussed a quantization of the electromagnetic field
relying on the solution of the functional Schr\"odinger equation.
We apply a method which using the ground state solution  of the
Schr\"odinger equation and a stochastic equation allows to derive
a general solution of the Schr\"odinger equation. We discussed the
particular solution of the form $\psi_{CS}=\exp(-W_{CS})$ where
$W_{CS}$ is the Chern-Simons action. We have shown that the
general solution of the Schr\"odinger equation with the initial
condition $\exp(-W_{CS})\chi$ can be expressed in the form
$\exp(-W_{CS})\chi_{t}$, where $\chi_{t}$ is an expectation value
over a stochastic self-duality equation. The method exhibits the
relevance of the Chern-Simons wave functions and self-duality
equations (related to electromagnetic fields with a definite
helicity) in    quantum electrodynamics. We have derived the
corresponding equations for non-Abelian gauge theories. In the
latter case a solution of the non-linear stochastic self-duality
equations can be obtained in perturbation expansion or by
numerical methods.

 Our
main motivation is a prospective application of the method to a
quantization of gravity. It is expected that quantum gravity is an
extension of a quantum theory of massless spin 2 tensor fields
(gravitons). We have applied the Chern-Simons wave function method
to gravitons in a close analogy to the quantization of the
(non-linear) gauge fields.  We have found the Chern-Simons wave
function in the linearized gravity. If such a wave function is
known (without the linear approximation) then the problem of
deriving a general solution of the Wheeler-deWitt equation can be
reduced to the problem of solving a non-linear self-duality-type
stochastic equation. This is still a work in progress.

\section{\bf Appendix}
The Chern-Simons states $\psi_{CS}$ are interesting objects in
 mathematical physics because (formal) correlation functions in these states are related
to Gauss linking number, Jones polynomials, and in physics, to
fractional statistics. However, these states are not square
integrable. The question arises of whether some regularizations
$\psi=\chi\psi_{CS}$ of these states have similar properties. We
are going to establish  some approximations expressed by the
formula (34)( where $U_{t}$ denotes the unitary evolution (19))
\begin{equation}\begin{array}{l}
(\psi_{CS}\chi,{\bf A}_{t}^{Q}({\bf x}){\bf A}^{Q}({\bf
y})\psi_{CS}\chi)(\psi,\psi)^{-1}=(U_{t}\psi_{CS}\chi,{\bf A}({\bf
x})U_{t}{\bf A}({\bf y})\psi_{CS}\chi)(\psi,\psi)^{-1}
 \cr\rightarrow -(2M)^{-1}\exp(-tM )({\bf x},{\bf
y})\end{array}
\end{equation} when $ \chi\rightarrow 1$.
We are unable to prove this limit for a general $\chi$ but we
supply a heuristic argument why the rhs of eq.(81) can  be a good
approximation for the lhs.  We rewrite the lhs of eq.(34) as
\begin{equation}\begin{array}{l}
 (\psi_{t},{\bf A}({\bf x})
E[{\bf A}_{t}({\bf y})]\psi_{CS})+ (\psi_{t},{\bf A}({\bf x})
E[{\bf A}_{t}({\bf y})(\chi({\bf A}_{t})-1)]\psi_{CS}).
\end{array}\end{equation}
From eq.(35)\begin{equation} E[{\bf A}_{t}({\bf
y})]=(\exp(-tM){\bf A})({\bf y}).
\end{equation}
Hence, if the second term in eq.(82) tends to zero when
$\chi-1\rightarrow 0$ and
\begin{equation}
(\psi_{t},A_{j}({\bf x})A_{k}({\bf y})\psi_{CS})\rightarrow
-(2M)^{-1}({\bf x},{\bf y})_{jk}\end{equation} when
$\psi_{t}=\psi_{CS}\chi_{t}\rightarrow \psi_{CS}$ for
$\chi\rightarrow 1$ then the limit (81) holds true. The limit (84)
relies on the formula
\begin{displaymath}
\vert\det M\vert^{\frac{1}{2}}\int d{\bf A}\exp({\bf A}M{\bf
A}){\bf A}{\bf A}=-\frac{1}{2}M^{-1}.
\end{displaymath}
which will be shown as a limit in eq.(86) below for a special
class of regularizations. It would be rather difficult to prove
eqs.(81)and (84) for general $\chi$. We show that these equations
hold true for a specific Gaussian regularization (we expect that
they could be proved for all Gaussian regularizations).

 In this Appendix we discuss
the following regularization
\begin{equation}\begin{array}{l}
\psi^{\alpha}=\psi_{CS}\chi=\psi_{CS}\exp\Big(-\frac{\alpha}{2\hbar}{\bf
A}\sqrt{-\triangle}{\bf
A}\Big)\cr=\exp\Big(-\frac{1}{2\hbar}\tilde{A}_{j}(\alpha\vert{\bf
p}\vert\delta_{jk}-i\epsilon_{jkl}p^{l})\tilde{A}_{k}\Big)\equiv
\exp\Big(-\frac{1}{2\hbar}\tilde{A}_{j}R_{jk}\tilde{A}_{k}\Big),
\end{array}\end{equation}where $\tilde{A}$ denotes the Fourier transform.
If $\Re\alpha>1$ then the state $\psi^{\alpha}$ is square
integrable. Using eq.(32) and the formula
$B^{-1}=\int_{0}^{\infty}ds\exp(-sB)$ (for a matrix with $\Re
B>0$) we can calculate
\begin{equation}\begin{array}{l}
(\psi^{\alpha},A_{j}({\bf x})A_{k}({\bf
y})\psi^{\alpha})(\psi^{\alpha},\psi^{\alpha})^{-1}=\frac{1}{2}\int
d{\bf p} (\frac{\alpha}{\alpha^{2}-1}\vert{\bf
p}\vert^{-1}(\delta_{jk}-p_{j}p_{k} {\bf
p}^{-2})\cr-\frac{i}{\alpha^{2}-1}\epsilon_{jkl}p_{l}\vert {\bf
p}\vert^{-2})(2\pi)^{-3}\exp(i{\bf p}({\bf x}-{\bf
y}))=R^{-1}({\bf x},{\bf y})_{jk}
\end{array}\end{equation}
for $\Re\alpha>1$. It is interesting to note that eq.(86) is
singular for $\alpha=1$ (as expected) but it is finite for other
values of $\alpha$. The term in the first line in eq.(86) is
vanishing when $\alpha\rightarrow 0$ whereas the second term
preserves its form just changing the sign in this limit (hence,
the change of sign on the rhs of eq.(84)). It follows that the
limits (81) and (84) hold true for states (85) at $t=0$.

For $t\geq 0$ we need to calculate $\psi_{t}$. We may consider the
electromagnetic field in a finite volume with periodic boundary
conditions. Then, in the momentum representation the state (85) is
a product state over momenta. When we restrict the number of
momenta then we have a finite product of Gaussian functions in the
product state (85). We calculate $\psi_{t}$ for a finite number
$N$ of factors and subsequently take the limit $N\rightarrow
\infty$.
 For the calculation of the time evolution we apply the evolution kernel $K_{t}$
of the free electromagnetic field
\begin{equation}
    \begin{array}{l}
        K_{t}({\bf A},{\bf A}^{\prime})=\Big(\det \frac{i\omega
        }{2\pi\hbar\sin(\omega t)}\Big)^{\frac{1}{2}} \cr \exp\Big( i{\bf
            A}\frac{\omega \cos(\omega t)}{2\hbar\sin(\omega t)}{\bf A}+i{\bf
            A}^{\prime}\frac{\omega \cos(\omega t)}{2\hbar\sin(\omega t)}{\bf A}^{\prime}
        -i{\bf A}\frac{\omega }{\hbar\sin(\omega t)}{\bf
            A}^{\prime}\Big),
\end{array}\end{equation}where $\omega=\sqrt{-\triangle}$.
In the momentum space the evolution kernel is a product of
Gaussian factors. First, we show the limit (84). For this purpose
we calculate $\psi_{t}$ explicitly. Let
\begin{equation}
{\cal R}_{jk}=\Big(\vert{\bf p}\vert(\alpha-i\cot(\vert{\bf
p}\vert t))\Big)\delta_{jk}-i\epsilon_{jkl}p^{l}.
\end{equation}We note that in eq.(88) we obtain ${\cal R}$
from $R$ in eq.(85)just by a replacement $\alpha\rightarrow
\alpha-i\cot(\vert{\bf p}\vert t)$ . For later purposes we
calculate an evolution of a more general state
\begin{equation}\begin{array}{l}
U_{t}\Big(\exp(\frac{i}{\hbar}\int d{\bf x}{\bf J}{\bf
A})\psi_{CS}\chi\Big) \cr=Z(t) \exp\Big(\Big(\frac{i}{2\hbar}{\bf
A}\omega\cot(\omega t){\bf A}-\frac{1}{2\hbar}({\bf
J}-\frac{\omega}{\sin(\omega t)}{\bf A}){\cal R}^{-1}({\bf
J}-\frac{\omega}{\sin(\omega t)}{\bf
A})\Big),\end{array}\end{equation} where $Z(t)$ is a numerical
factor which cancels in the final result.

We  note that ${\cal R}^{-1}$ is given by eq.(86) where
$\alpha\rightarrow\alpha-i\cot(\vert{\bf p}\vert t)$. We
considered the state (89) because we can obtain ${\bf A}({\bf y})$
by differentiation $i{\bf A}({\bf y})=\frac{\delta}{\delta {\bf
J}({\bf y})}$ at ${\bf J}=0$. It follows that the first term in
eq.(82) is
\begin{equation}\begin{array}{l}
(\psi_{t},{\bf A}({\bf x}) E[{\bf A}_{t}({\bf y})]\psi_{CS})\cr=
\frac{1}{i}{\cal R}^{-1}\frac{\omega}{\sin(\omega t)}
\Big(-\frac{\omega}{\sin(\omega t)}{\cal
R}^{-1}\frac{\omega}{\sin(\omega t)} +i\omega \cot(\omega
t)\Big)^{-1}({\bf x},{\bf y}).
\end{array}\end{equation}
We can calculate this expression in the momentum space with ${\cal
R}^{-1}$ from eq.(86)($\alpha\rightarrow \alpha-i\cot(\vert{\bf
p}\vert t)$). We derive from eq.(90) the rhs of eq.(84) expressed
in the momentum space as
\begin{equation}\begin{array}{l}
        \Big(-(2M)^{-1}\exp(-tM)\Big)_{jk}\cr
        =\frac{1}{2}\Big((\delta_{jk}-p_{j}p_{k} {\bf p}^{-2})\vert {\bf
            p}\vert^{-1}\sin(t\vert{\bf p}\vert)+\epsilon_{jkl}p_{l}\vert
        {\bf p}\vert^{-2}(\cos(t\vert{\bf p}\vert)\Big)
    \end{array}
\end{equation}
as follows from the formula \begin{displaymath}
 \Big(M^{-1}\exp(-tM)=-\int_{0}^{t}ds\exp(-sM)+M^{-1}
\end{displaymath}and eq.(32). The second term in eq.(82) can be
calculated by means of eq.(89) as
\begin{displaymath}
(U_{t}\psi_{CS}\chi,{\bf A}({\bf x})U_{t}{\bf A}({\bf
y})\psi_{CS}(\chi-1))
\end{displaymath} and easily estimated as tending to zero.
 We did these calculations in
order to show how the heuristics (82) works.However, with the
explicit formula for  $\chi$ using eqs. (86) and (89) we can
directly calculate (81) with an application of known formulas for
Gaussian integration. Then, knowing ${\cal R}^{-1}$ from eq.(86)
we can follow the dependence of eq.(81) on $\alpha$ leading in the
limit $\alpha\rightarrow 0$ to the result on the rhs of eq.(81).

 {\bf Data
Availability Statement}:No Data associated in the manuscript.

  \end{document}